\documentclass [titlepage,letterpaper,11pt] {article}

\usepackage{url}
\usepackage[spanish, english]{babel}
\usepackage[T1]{fontenc}
\usepackage[latin1]{inputenc}
\usepackage[pdftex]{color,graphicx}
\usepackage{fullpage}
\usepackage{graphicx}
\usepackage{setspace}

\author{
  Priscill Orue-Esquivel\footnote{Corresponding author. Phone number: $+595-972-646131$}, Bartolomé Rubio\\
  \texttt{Languages and Computer Science Dept.} \\
  \texttt{University of Málaga} \\
  \texttt{Bulevar Louis Pasteur, 35. 29071} \\
  \texttt{Málaga, Spain} \\
  \texttt{$priscill\_orue@alu.uma.es$, $tolo@lcc.uma.es$} \\
  }

\title{WiSANCloud: a set of UML-based specifications for the integration of Wireless Sensor and Actor Networks (WSANs) with the Cloud Computing}

\begin{document}

\maketitle

\begin{abstract}
Giving the current trend to combine the advantages of Wireless Sensor and Actor Networks (WSANs) with the Cloud Computing technology, this work proposes a set of specifications, based on the Unified Modeling Language - UML, in order to provide the general framework for the design of the integration of said components. One of the keys of the integration is the architecture of the WSAN, due to its structural relationship with the Cloud in the definition of the combination. Regarding the standard applied in the integration, UML and its subset, Systems Modeling Language - SysML , are proposed by the Object Management  Group - OMG\textsuperscript{\textregistered} to deal with cloud applications; so, this indicates the starting point of the process of the design of specifications for WSAN-Cloud integration. Based on the current state of UML tools for analysis and design, there are several aspects to take into account in order to define the integration process.

\end{abstract}

\section{Introduction}
	Cloud Computing contains applications ``exposed as sophisticated services that can be accessed over a network'' \cite{buyya:2009}. For this reason, its advantages are used in combination with other automated systems like WSANs (Wireless Sensor and Actor Networks). The integration of WSANs with the Cloud Computing has gained attention due to the benefits of their structural relation. The Cloud provides scalability in terms of processing power and different types of interconnected services while WSANs contain various nodes in charge of capturing and performing a local pre-processing of data.

	Based on the review of the state-of-the-art about the WSAN-Cloud Computing integration, there is not a definite model for the processes of analysis and design of the said integration. One of the reasons is because the Cloud per se lacks standards for the interoperability among service providers, consumers and developers \cite{blaisdell:2011}.
		
	Of the various paradigms and approaches for analysis and design of systems, the Unified Modeling Language (UML) is one of the most studied ones for the Cloud environment, as Drusinsky et al \cite{drusinsky:2011} and Kurschl and Beer \cite{kurschl:2009} show in their research. As a result, the main motivation of this work is to contribute a set of specifications that might be applicable to the integration of WSANs with the Cloud Computing, using UML as a basic standard.
	
	As an initial step, Cloud Computing and WSANs are studied separately from the perspective of the processes of analysis and design of applications. Following, there is a reflection upon the selection of UML as the basic standard of this work. As a final section of the theoretical framework, there is a step-by-step analysis demonstrating the lag of the software engineering process of WSAN-Cloud integration. For the contribution section, there are separate studies of how the Cloud and the WSANs focus the processes of analysis and design. Next, the relevant aspects of each component are taken and combined to propose the specifications for the integration. As a practical demonstration, an example of integration design process is performed. Finally, several conclusions related to the work - considering the methods and the tools - are presented at the end of this report.


\section{Theoretical Fundamentals and Related Work}	
		This section is intended to offer a review about the theoretical basis of the work. It is divided into three sub-sections. It starts with the descriptions of the situation of Cloud Computing and the Wireless Sensor and Actor Networks (WSANs, in order to define the aspects related to engineering processes (analysis, design and development) within the Cloud. As the final part of this section, the Unified Modeling Language (UML) is studied to identify the elements that could be applied in the set of specifications for WSANs-Cloud Computing integration that is going to be defined afterwards.
		
	\subsection{Cloud Computing (CC): application development state-of-the-art}
	Cloud Computing is an emerging technology that provides ``on-demand computational capacity as a service'' \cite{tao:2011}. This section aims to describe the state-of-the-art of how applications are developed with this approach. First, the processes of analysis, design and development within the Cloud are analyzed. Then, a brief discussion of the current state and challenges for the development of Cloud applications is included.

	\subsubsection{Analysis, design and development of applications within the Cloud}
	The Cloud Standards Wiki \cite{cswiki:2012} comprises the efforts to standardize the work related to Cloud Computing. There are several institutions contributing with standards from different areas within the Cloud. These institutions and their areas of work are summarized in the Table \ref{tab:cloudstandards}.
	
	\begin{table}[!hbp]
	\centering
	\begin{tabular}{|p{7cm}|p{10cm}|}
	\hline
	\textbf{Organization} & \textbf{Contribution}  \\
	\hline
	Cloud Security Alliance & Promotes ``the use of best practices for providing security assurance within Cloud Computing, and provide education on the uses of Cloud Computing to help secure all other forms of computing'' \\
	\hline
	Cloud Standards Customer Council & End user advocacy group that is dedicated to accelerate the ``cloud's successful adoption, and drilling down into the standards, security and interoperability issues surrounding the transition to the cloud.'' \\
	\hline
	Distributed Management Task Force (DMTF) & Focused on ``standardizing interactions between cloud environments by developing cloud management use cases, architectures and interactions.'' \\
	\hline
	European Telecommunications Standards Institute (ETSI) & Aims to ``address issues associated with the convergence between IT (Information Technology) and Telecommunications'' \\
	\hline
	National Institute of Standards and Technology (NIST) & Working on the definition of Cloud Computing \\
	\hline
	Open Grid Forum (OGF) & A organization that develops standards ``operating in the areas of grid, cloud and related forms of advanced distributed computing.'' \\
	\hline
	Object Management Group (OMG)\textsuperscript{\textregistered} & Focuses ``on modeling, and the first specific cloud-related specification efforts have only just begun, focusing on modeling deployment of applications and services on clouds for portability, interoperability and reuse.'' The most important specification related to the Cloud is SysML, which will be later discussed in this work. \\
	\hline
	Open Cloud Consortium (OCC) & It has a particular focus in large data clouds \\
	\hline	
	Organization for the Advancement of Structured Information Standards (OASIS) & Leads the ``development, convergence and adoption of open standards for the global information society.'' Most of its foundational standards are a natural extension of SOA and network management models. \\
	\hline
	Storage Networking Industry Association (SNIA) & Originated ``the Cloud Storage Technical Work Group for the purpose of developing SNIA Architecture related to system implementations of Cloud Storage technology'' \\
	\hline
	Cloud Work Group & Focuses on creating ``a common understanding among buyers and suppliers of how enterprises of all sizes and scales of operation can include Cloud Computing technology in a safe and secure way in their architectures to realize its significant cost, scalability and agility benefits'' \\
	\hline	
	Association for Retail Technology Standards (ARTS) & They work on the usage of the Cloud for retailers \\
	\hline
	TM Forum & Helps enterprises to adopt and use digital information services for business effectiveness \\
	\hline				
	\end{tabular}
	\caption{Summary of organizations working on Cloud standards \cite{cswiki:2012}}
	\label{tab:cloudstandards}
\end{table}

		The most relevant effort for this work is the one from the Object Management Group (OMG)\textsuperscript{\textregistered}. OMG\textsuperscript{\textregistered} offers SysML -a subset of UML- as a tool for modeling within the Cloud. According to the OMG\textsuperscript{\textregistered}, SysML ``is a general-purpose graphical modeling language for specifying, analyzing, designing, and verifying complex systems that may include hardware, software, information, personnel, procedures, and facilities'' \cite{omg:2011}. Therefore, this tool appears to be a promising one that would help to reach the goals of this work. 
		
		However, when reviewing the SysML specifications, there is no direct relationship with the Cloud. This implies that The Cloud Standards Wiki \cite{cswiki:2012} contemplates SysML as an alternative for analysis and design in the Cloud context, but the modeling tool is not directly related to the Cloud. Thus, there is a unidirectional relationship. In brief, the processes of analysis, design and development of applications within the Cloud are not overtly defined in terms of specifications or standards, but there is a tentative standard that can be reused for the purposes of this work.

	\subsubsection{Current state and its challenges for application development}

\begin{figure}[h]
	\centering
		\includegraphics{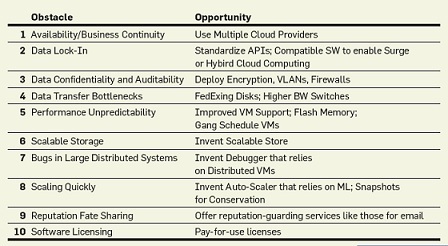}
	\caption{Obstacles and opportunities for Cloud Computing growth \cite{armbrust:2010}}
	\label{fig:obstacles}
	\end{figure}

	Armbrust et al \cite{armbrust:2010} summarizes the current state and the challenges offered by the Cloud Computing in Figure \ref{fig:obstacles}. The first obstacle-opportunity can be handled by having multiple Cloud Computing providers. ``One solution would be to standardize the APIs in such a way that a SaaS developer could deploy services and data across multiple cloud computing providers so that the failure of a single company would not take all copies of customer data with it.'' On the other hand, the responsibility for security is shared by many parties; and developers have to check their applications comply with security standards \cite{armbrust:2010}.

	Another of the functional requirements of a Cloud application is the scalability in storage. The idea is ``to create a storage system that would not only meet existing programmer expectations in regard to durability, high availability, and the ability to manage and query data, but combine them with the cloud advantages of scaling arbitrarily up and down on demand.'' Virtual Machines constitute a plausible solution to remove errors and bugs in a large system like the Cloud. It is important to bear in mind that applications ought to respect the Service Level Agreements - SLAs. This is known as dynamic scaling. Security issues like levels of access and authorization are important in an area where all users share resources. Finally, software licensing is an issue that is part of the analysis process of application development in the Cloud \cite{armbrust:2010}.

	Tao et al \cite{tao:2011} claims that ``Cloud technology is currently still in the development phase. Many issues, like data security and monitoring, are still not considered or have to be improved.'' For these reasons, plus the ones described by Armbrust \cite{armbrust:2010}, the current situation of the Cloud is defined by the presence of various standards aiming to reach a consensus on several aspects. The main challenge, however, is to define a specific standard for application development.
	
	\subsection{Wireless Sensor and Actor Networks (WSANs): current situation of application development}
	Wireless Sensor and Actor -sometimes referred to as ``actuators'' in the literature- Networks are capable of ``observing the physical world, processing the data, making decisions based on the observations and performing appropriate actions''. This section introduces several concepts related to WSANs. In the beginning, a comprehensive analysis of the structure and operation of WSANs is performed. Then, the contexts of WSANs' applications are presented, with a strong emphasis with Cloud Computing integration. Finally, there is a description of the current situation of the processes of analysis, design and development of applications with WSANs.

		\subsubsection{Operation and structure of WSANs}
\begin{figure}[h]
	\centering
		\includegraphics{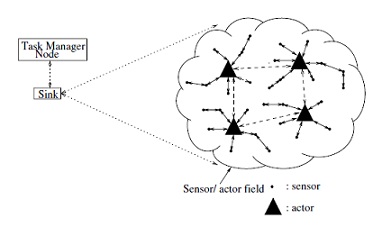}
	\caption{The physical architecture of WSANs \cite{akyildiz:2004}}
	\label{fig:wsansarchitecture}
\end{figure}

		There are two important roles in a WSAN: sensor and actor. A WSAN collects data from the environment where it is inserted and then performs appropriate actions taking into account the collected data. Like Figure \ref{fig:wsansarchitecture} shows, ``these nodes are scattered in the sensor/actor field while the sink monitors the overall network and communicates with the task manager node and sensor/actor nodes'' \cite{akyildiz:2004}. 

\begin{figure}[!hbp]			
	\centering
		\includegraphics{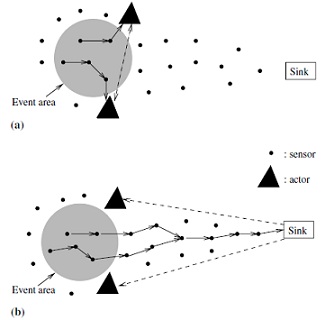}
	\caption{Automated and Semi-automated architecture of WSANs \cite{akyildiz:2004}}
	\label{fig:architectures}
\end{figure}

	Sensors that detect a phenomenon can transmit their data to the actor nodes which process all incoming data and initiate appropriate actions; this is called Automated Architecture because of the absence of a central controller, like human interaction. Another alternative is when sensors route data back to the sink which issues action commands to the actors in the network. In this case, the sink - acting as the central controller - collects data and coordinates the acting process \cite{akyildiz:2004}. The difference is perceived in Figure \ref{fig:architectures}	

	Sensor nodes contain a power unit, communication subsystems (receiver and transmitter), storage and processing units, Analog to Digital Converter (ADC) and the sensing unit. The role of the sensing unit is to monitor phenomena like thermal, optic or acoustic events. Then, ``the collected analog data are converted to digital data by ADC and then are analyzed by a processor and then transmitted to nearby actors'' \cite{vandih:2005}. 
	
\begin{figure}[h]
	\centering
		\includegraphics{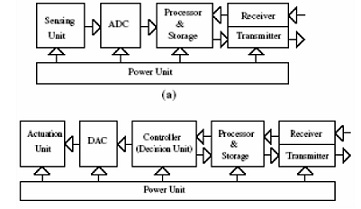}
	\caption{(a) Sensor and (b) Actor nodes architectures \cite{vandih:2005}}
	\label{fig:SAarchitectures}
\end{figure}

	The controller, which is a decision unit, takes sensor readings as input and produces action commands as output. At that point, the commands are converted to analog signals by the Digital to Analog Converter (DAC) and are transformed into actions ``via the actuation units''. There are integrated sensor/actor nodes that can substitute actor nodes. ``Since an integrated sensor/actor node is capable of both sensing and acting, it has sensing unit and ADC in addition to all components of an actor node.'' \cite{vandih:2005}. Figure \ref{fig:SAarchitectures} displays the structure of both sensors and actors.	

	\subsubsection{Cloud contexts of WSANs' application}
	The use of Cloud Computing as a platform for the operations of WSANs presents advantages. Cloud Computing offers the potential of increasing capacity and extending other features of capability while the system is under operation, without adding new infrastructure, training new human resources or licensing software. For these reasons, the integration of WSANs with Cloud Computing is a field in development with promising results. Kurschl and Beer \cite{kurschl:2009} suggest several aspects to take into account when dealing with WSAN-Cloud integration. These aspects consist of various basic services. 
	
	Regarding examples of WSAN-Cloud integration, Kurschl and Beer \cite{kurschl:2009} propose a system in which they combine two architecture models: ``1. SOA Roles in proposed Architecture; and 2. Internet and Integration Controller interaction Architecture(IICiA) (using cloud technology)''. The sensor nodes and integration controller are able to interact through SOA architecture. Sensor nodes constitute service providers and sink nodes as seen as consumers the get the information through the Integration Controller.
	
\begin{figure}[h]
	\centering
		\includegraphics{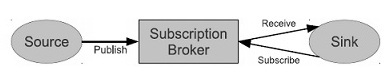}
	\caption{Architecture of Publish/Subscribe systems \cite{walters:2011}}
	\label{fig:publishsubscribe}
\end{figure}

	Another example of WSAN-Cloud integration is through the Publish-Subscribe systems, proposed by Walters \cite{walters:2011}. These systems offer a mechanism to disseminate information where there is no need of a priori knowledge regarding information requirements. Figure \ref{fig:publishsubscribe} shows an information source enabling an interface for sinks to submit a subscription request. When the request is approved, the ``source periodically sends updated information to the sink based on the subscription request''. The source owner has no need to know this in advance of the request. One point in detriment of these systems is their application in large infrastructures due to the need of resources to handle large numbers of requests. However, there are some approaches in \cite{frey:2007} trying to solve this disadvantage.
	
	\subsubsection{Analysis, design and development of applications with WSANs}
	The problem of the lack of consensus for the development of WSAN applications goes back to WSNs (Wireless Sensor Networks), the predecessors of WSANs. Zhao and Guibas \cite{zhao:2004} affirm that new technological advances bring new challenges for ``information processing in sensor networks. What is needed are novel computational representations, algorithms and protocols and design methodologies and tools  to support distributed signal processing, information storage and management, networking, and application development.''
	
	However, several efforts have been initiated towards a model of development of WSAN applications. ``The provisioning of sensor information is seen as a basic requirement for enabling information and services that respond to changes in context of state or being''. Current and existing research has depended mainly on the principles of middleware solutions, which allows acquisition and dissemination of useful information \cite{walters:2011}. 
	
	From 	Kurschl and Beer's experience \cite{kurschl:2009}, the current available tools for design and analysis have advantages. Several programming languages and different kind of devices can be put in service and easily interconnected. ``It provides a lot of flexibility in terms of programming languages and devices, but this come with the drawback of heterogeneity, which quite often leads to more complexity.''.
	
	In brief, there are models and strategies to perform the design and development of WSAN-based applications. From the reviewed literature, the most remarkable one is the use of middlewares. Nevertheless, evidence of models or techniques for analysis was not found. 

	\subsection{Unified Modeling Language (UML): concepts for WSAN-Cloud integration}
	``The major benefits of OO modeling, design and subsequent implementation are reusability, reliability, robustness, extensibility, and maintainability'' and these features are essential to keep on with WSAN-Cloud application development. UML is accepted in many software development organizations as a tool to standardize processes. Within the Cloud context, where application development evolves in a fast pace, it is necessary to rely on approaches that emphasize re-usability and encapsulation. The main idea is to build models which constitute an architecture ``with details hidden inside shared objects exposing interfaces and methods for interaction and usage''\cite{walters:2011}.
		
	Objects can be serialized, which is to convert them into bits for storage in file \cite{greene:2011}. This feature makes possible that linked architectures can construct and deliver objects where requested, protecting sensitive information, like sensor passwords. ``Such a model enables a high level of dynamic behavior; and when coupled with an object oriented persistence,'' there is a responsive and scalable solution \cite{walters:2011}. 

	\subsection{Common aspects from WSANs and Cloud Computing related to analysis, design and development of applications}
	This is a summary of the processes of analysis, design and development from three viewpoints: Cloud Computing, WSANs and UML. The idea is to put in a nutshell the concepts studied in the theoretical framework, in order to have a view of the work which will be developed.
	
	\begin{table}[!hbp]
	\centering
	\begin{tabular}{|p{2cm}|p{4cm}|p{4cm}|p{6cm}|}
	\hline
	\textbf{Process / Area} & \textbf{Cloud Computing} & \textbf{WSANs} & \textbf{UML}  \\
	\hline
	Analysis & OMG proposes SysML as an alternative, according to The Cloud Standards Wiki \cite{cswiki:2012} & No proposed standards or tools in reviewed literature & The most approximate tool for the purposes of this work is SysML as a ``a general-purpose graphical modeling language for specifying, analyzing, designing, and verifying complex systems that may include hardware, software, information, personnel, procedures, and facilities'' \cite{omg:2011}. \\
	\hline
	Design & OMG proposes SysML as an alternative, according to The Cloud Standards Wiki \cite{cswiki:2012} & No proposed standards or tools in reviewed literature & The most approximate tool for the purposes of this work is SysML as a ``a general-purpose graphical modeling language for specifying, analyzing, designing, and verifying complex systems that may include hardware, software, information, personnel, procedures, and facilities'' \cite{omg:2011}. \\
	\hline
	Development & The Distributed Management Task Force (DMTF) is working on issues like portability, cloud architecture, interfaces, management and auditing \cite{cswiki:2012} & Commonly, Middlewares are used as the basis for application development (Walters \cite{walters:2011}) & Not applicable \\
	\hline
	\end{tabular}
	\caption{Summary of the processes in application development based on WSAN-Cloud Integration}
	\label{tab:summmaryprocesses}
\end{table}
	
	From the summary presented in Table \ref{tab:summmaryprocesses}, there is a lag in the software engineering process applied in WSANs and a proposal of a framework of analysis and design from Cloud Computing, based on UML. Therefore, the main goal of this work is to create a link of analysis and design of applications based on the WSANs-Cloud integration, through the UML standard; thus, the concept of reutilization will be applied, to avoid ``reinventing the wheel''.

	\section{Contributions: design of UML-based specifications for the integration of WSANs with CC }
	This section is the practical contribution of the work. The goal is to offer a set of specifications for the integration of WSANs - Wireless Sensor and Actor Networks with Cloud Computing, using the UML as the basic standard. From the theoretical point of view, the main tool will be SysML, which has already been proposed by the OMG for Cloud Computing and the other aspect to take into account will be the processes of analysis and design, which are inexistent in WSAN development.

	\subsection{Design of specifications to integrate WSANs with the CC, considering UML}
	The design of specifications to integrate WSANs with the Cloud technology takes into account UML as a basic standard for this work. The aspects to be considered are taken from the generic UML and a subset of it called SysML (System Modeling Language). ``SysML reuses a subset of UML 2 and provides additional extensions to satisfy the requirements of the language'' \cite{omg:2011}
				
	Therefore, this section will analyze the requirements for WSANs and Cloud Computing, both separately and together. The steps to be followed consist in the following:
\begin{itemize}
	\item Define and describe analysis and design-related aspects about WSANs
	\item Define and describe analysis and design-related aspects about the Cloud
	\item Define and describe analysis and design-related aspects from WSANs and the Cloud together
	\item Define and propose a set of specifications using UML/SysML.
\end{itemize}
		
	\subsubsection{Analysis and design-related aspects about the Cloud}
\begin{figure}[!hbp]
	\centering
		\includegraphics{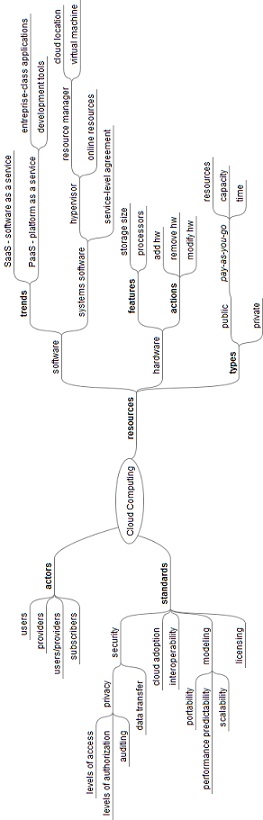}
	\caption{Cloud Computing Overview}
	\label{fig:ccoverview}
\end{figure}

	Figure \ref{fig:ccoverview} shows an overview of the Cloud aspects to consider when developing an application, both for SaaS and PaaS. As it can be seen, the nodes directly involved with the Cloud technology are: actors, resources and standards. This means that it is crucial to analyze the entities that are going to play roles in the infrastructure, the tools that are going to be used including which target the work will be aimed at and the points in question to deal with usability and maintainability.	

\begin{figure}[!hbp]
	\centering
		\includegraphics{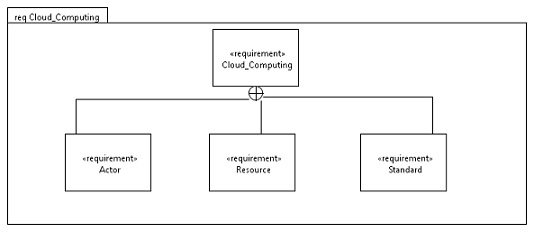}
	\caption{Cloud Computing Requirement Diagram}
	\label{fig:CloudRD}
\end{figure}

	After the general overview of the aspects to deal within the Cloud, Figure \ref{fig:CloudRD} represents the categories to be studied in the form of a Requirement Diagram. Being the Cloud a starting point, the whole structure is depicted in a hierarchical form. ``A composite requirement can contain subrequirements in terms of a requirements hierarchy, specified using the UML namespace containment mechanism.'' \cite{omg:2011}

\begin{figure}[!hbp]
	\centering
		\includegraphics{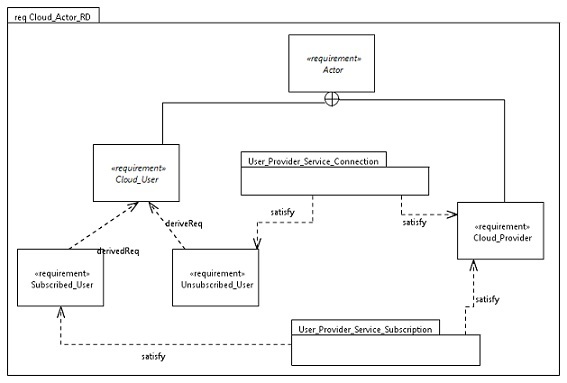}
	\caption{Actors of the Cloud in a Requirement Diagram}
	\label{fig:CloudActorRD}
\end{figure}

	Figure \ref{fig:CloudActorRD} includes all the actors in the Cloud. Due to limitations of the modeling tool \cite{eclipsebugs:2011}, it is important to note that blocks containing the link between users, providers and services are represented as packages that satisfy the requirements. Moreover, it is denoted that clients might not need a subscription to access the provided services, but it is still suggested that a temporary type of subscription be created while the provision of the service is taking place.
	
\begin{figure}[h]
	\centering
		\includegraphics{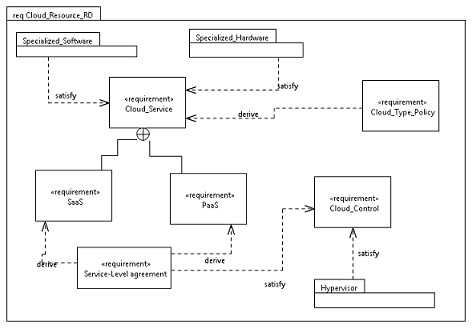}
	\caption{Cloud Resource in a Requirement Diagram}
	\label{fig:CloudResourceRD}
\end{figure}

	Regarding resources, Figure \ref{fig:CloudResourceRD} details the requirements to build a Cloud. As mentioned in the theoretical section, this work focuses in the Cloud from two viewpoints: Software-as-a-Service (SaaS), and Platform-as-a-Service (PaaS); this is the reason why the \emph{$Cloud\_Service$} requirement is decomposed in two other requirements. From those points of view, both, the \emph{$Service\_Level\_Agreement$} and \emph{$Cloud\_Type\_Policy$} requirements are requirements that need to satisfy the two approaches, in order to have them working properly.
	
\begin{figure}[h]
	\centering
		\includegraphics{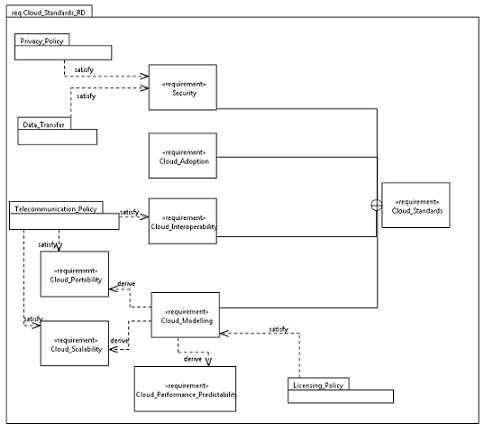}
	\caption{Cloud Standards represented in a Requirement Diagram}
	\label{fig:CloudStandardsRD}
\end{figure}

 Taking into account all the efforts to standardize the Cloud, which are summarized in Table \ref{tab:cloudstandards}, Figure \ref{fig:CloudStandardsRD} represents the general requirements to operate, model and maintain Cloud applications. Packages represent external entities that satisfy the needs. For example, the package \emph{$Telecommunication\_Policy$} ensures that a given cloud is interoperable and portable with other ones, as well as it guarantees scalability of the modeled applications.

	\subsubsection{Analysis and design-related aspects about WSANs}
\begin{figure}[!hbp]
	\centering
		\includegraphics{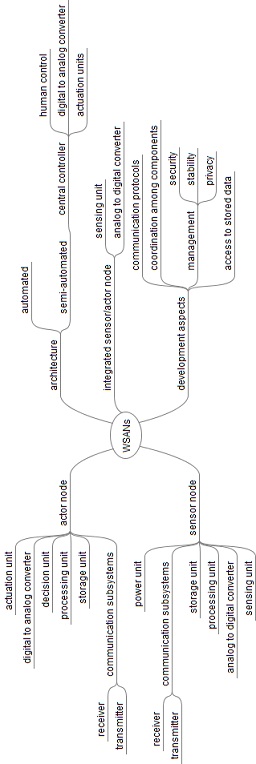}
	\caption{Requirements for WSANs}
	\label{fig:WSANsRequirements}
\end{figure}

	After identifying the requirements for Cloud-based applications, this section describes the requirements for Wireless Sensor and Actor Networks (WSANs). Figure \ref{fig:WSANsRequirements} denotes a general overview of WSANs' requirements, both from the structural and developmental points of view. The network architecture plays an important role when defining the pieces of the whole picture. This means that the branch \emph{development aspects} changes according to the structure of the network. However, for this work, all possible sensor-actor-controllerNode combinations are taken into account.
	
\begin{figure}[!hbp]
	\centering
		\includegraphics{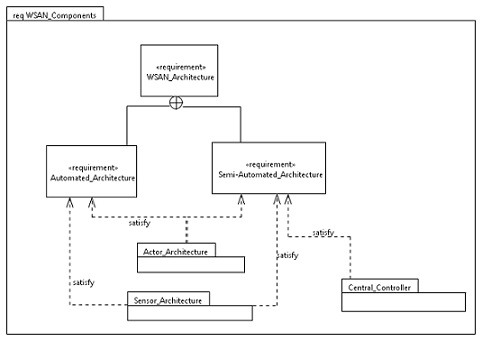}
	\caption{Requirements for WSAN Components}
	\label{fig:WSANComponents}
\end{figure}

From the map displayed in Figure \ref{fig:WSANsRequirements}, the corresponding analysis is as follows: structure and implementation. Figure \ref{fig:WSANComponents} details the components in terms of requirements and how they are associated. For the purposes of this analysis, the actor and sensor nodes are represented as packages, since they contain their own specifications for their implementation.

\begin{figure}[!hbp]
	\centering
		\includegraphics{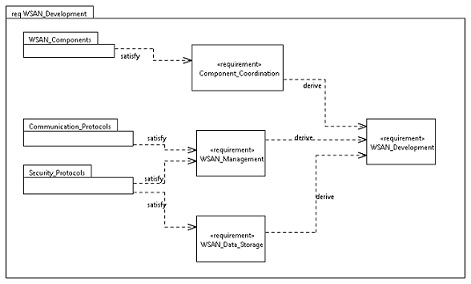}
	\caption{Requirements for WSAN Development}
	\label{fig:WSANDevelopment}
\end{figure}

Based on the standards listed in Table \ref{tab:cloudstandards}, node interconnection within the network is the first point in common when comparing with the components defined in the Cloud. Telecommunication aspects also play an important role to satisfy communication and storage needs.  These aspects plus the requirements defined in Figures \ref{fig:WSANComponents} and \ref{fig:WSANDevelopment} are taken into account for the development of WSAN-based applications. 

	\subsubsection{Analysis and design-related aspects from WSANs and the Cloud together}
	The analysis and definition of design-related aspects of the Cloud and WSANs are used to describe the points to deal when trying to integrate these two fields. As mentioned earlier, the first point in common between the Cloud and the WSANs refers to communication among internal entities and also to the external world. Therefore, the goal is to define the set of requirements to integrate WSANs to the Cloud.

\begin{figure}[h]
	\centering
		\includegraphics{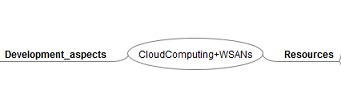}
	\caption{CloudComputing and WSANs proposed aspects for integration}
	\label{fig:CloudComputing+WSANs(1)}
\end{figure}

	Figure \ref{fig:CloudComputing+WSANs(1)} is a synthesis of the items enumerated as requirements in previous analysis of the Cloud and WSANs, separately. Since both entities share several attributes, the map is reorganized taking into account two main branches: \emph{$Development\_aspects$} and \emph{Resources}. In this way, what is available and what rules have to be fulfilled are displayed as requirements.


\begin{figure}[h]
	\centering
		\includegraphics{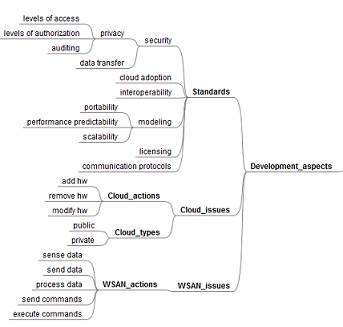}
	\caption{CloudComputing and WSANs: Development aspects}
	\label{fig:CloudComputing+WSANs(2)}
\end{figure}

	The \emph{$Development\_aspects$} detailed in figure \ref{fig:CloudComputing+WSANs(2)} takes into account the standards already proposed by organizations, cloud issues that include actions and types related to the Cloud, and actions performed by WSANs. Therefore, these aspects study the Cloud and the WSANs separately.

\begin{figure}[h]
	\centering
		\includegraphics{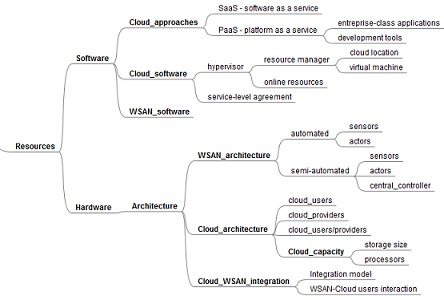}
	\caption{CloudComputing and WSANs: Resources}
	\label{fig:CloudComputing+WSANs(3)}
\end{figure}

	Regarding resources, figure \ref{fig:CloudComputing+WSANs(3)} sketches a reorganization of all the means employed both in the Cloud and the WSANs. There are two main types of resources: software and hardware. Here, the relevant information deals with the WSAN architecture: depending on the architecture assumed for the integration, the relation of Cloud roles with WSAN roles changes significantly. For this reason, two proposals of integration will be presented: one with automatic WSANs and the other with semi-automatic WSANs. As the branch \texttt{$Cloud\_WSAN\_Integration$} denotes, the integration can be carried out considering an integration model of the components and a WSAN - Cloud users interaction, which defines the way users obtain data from the integration. 

	\subsubsection{Proposal of a set of specifications using UML/SysML}
	After the preliminary analysis and definition of requirements for the integration of WSANs to the Cloud, this section goes deeper and proposes a set of specifications for the said integration. First, a Requirement Diagram is included to define the needs to satisfy in order to successfully integrate WSANs into the Cloud. Then, both UML and SysML are going to be applied to schematize the identified needs. Finally, a textual description will synthesize all specifications expressed in the form of diagrams. 
	
\begin{figure}[h]
	\centering
		\includegraphics{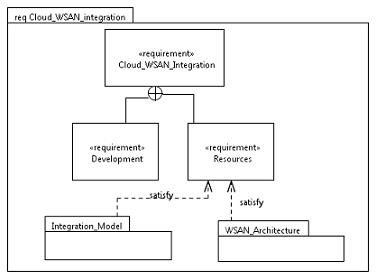}
	\caption{Cloud Computing and WSANs: synthesis}
	\label{fig:CloudWSAN}
\end{figure}

	The whole architecture of requirements can be synthesized like Figure \ref{fig:CloudWSAN} shows. From the \emph{Resources} side, the main aspects to consider are the \emph{Integration\_Model} because according to the selected model, the distribution of data and commands is defined; then, the \emph{WSAN\_Architecture} plays an important role due to the fact that provides the ways for the defined distribution of the information. In view of the existence of two types of WSAN architecture (automatic and semi-automatic), two sets of specifications for the integration to the Cloud will be presented.

	\textbf{a. Set of specifications for the integration of semi-automatic WSANs with the Cloud}: Since the Cloud has the roles of users and providers, WSANs - with their sensors and actors - fit in this overview. In other words, \emph{$Cloud\_users$} are related to \emph{WSAN actor nodes}, while \emph{$Cloud\_providers$} are related to \emph{WSAN sensor nodes}. On the other hand, Cloud capabilities can be applied to manage WSANs through a Publish/Subscribe model of integration.
	
\begin{figure}[h]
	\centering
		\includegraphics{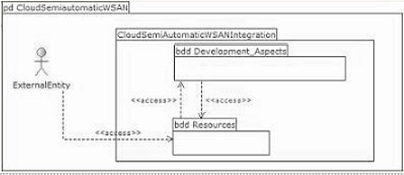}
	\caption{SemiAutomatic WSAN - Cloud Integration }
	\label{fig:semiautomaticintegration}
\end{figure}

	Necessary conditions for WSAN-Cloud integration are already defined via Requirement Diagrams; therefore, Block Diagrams are going to be used to propose the set of specifications. Blocks constitute modular units of system description and define a collection of features to describe a system under study. They may include ``both structural and behavioral features, such as properties and operations'' \cite{omg:2011}. To define a starting point, Figure \ref{fig:semiautomaticintegration} summarizes the areas to be worked on represented by packages, and how the system is related to external entities (users, other systems, etc) through the \emph{Resources} of the integration.
		
\begin{figure}[h]
	\centering
		\includegraphics{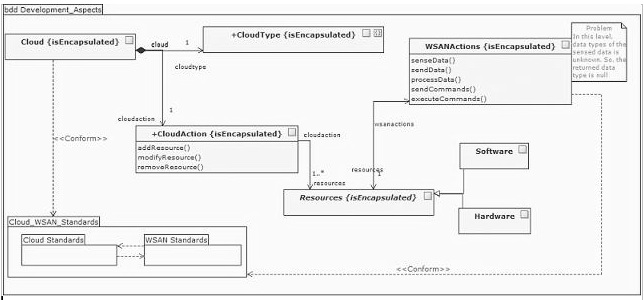}
	\caption{Block Diagram for the Development Aspects}
	\label{fig:bdddevelopmentaspects}
\end{figure}

	\emph{$Development\_aspects$} includes WSAN and Cloud elements, actions and standards. Regarding WSAN actions, data type of the sensed and processed data is unknown in this level (here, WSANs are treated generically), for this reason there are not data types specified in the Block Diagram of Figure \ref{fig:bdddevelopmentaspects}. Besides, standards are considered as packages that are conformed both by the Cloud and the WSAN to be integrated. Finally, since the two parties are independent in this level, all components are encapsulated.

\begin{figure}[h]
	\centering
		\includegraphics{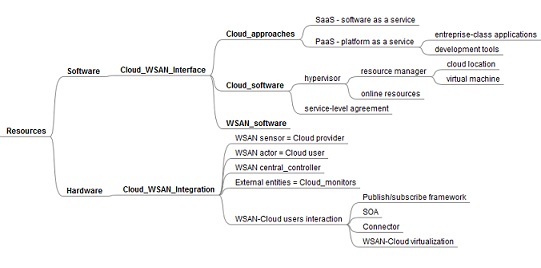}
	\caption{Resources for Cloud and semi-automatic WSAN integration}
	\label{fig:cloudsemiautomaticwsanintegration}
\end{figure}

	Figure \ref{fig:cloudsemiautomaticwsanintegration} displays the necessary resources to take into account when performing the integration. As mentioned earlier, the architecture of the integration relates WSAN sensors as Cloud providers, WSAN actors as Cloud users, External entities as Cloud monitors and the WSAN central controller that acts as the sink in the Publish/Subscribe model. Cloud monitors establish the criteria of how actors and sensors should perform and to which entities report the sensed data, for this reason, they are considered as external entities regarding the Cloud. 

\begin{figure}[h]
	\centering
		\includegraphics{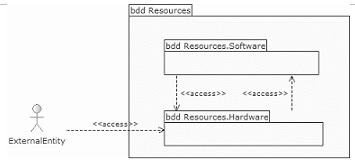}
	\caption{Overview of the resources for Cloud and semi-automatic WSAN integration}
	\label{fig:resourcesoverview}
\end{figure}

	The overview of the elements conforming the Resources is depicted in Figure \ref{fig:resourcesoverview}. In this level, the resources necessary for the integration are hardware and software. The \emph{ExternalEntity} actor, which maybe a human user or any other monitoring system, is directly related to the hardware because it deals with the architecture of the integration.

\begin{figure}[!hbp]
	\centering
		\includegraphics{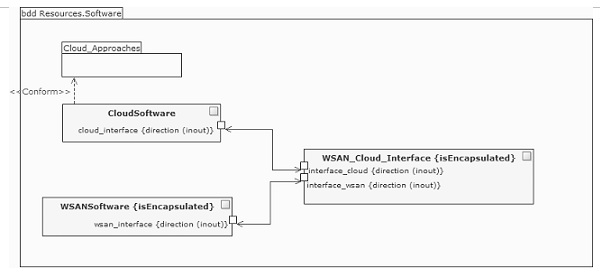}
	\caption{Block diagram about software resources (semi-automatic WSANs with the Cloud)}
	\label{fig:softwareresources}
\end{figure}

	The software resources needed for Semi-Automatic WSAN and Cloud integration are the same ones related to both parties, to ensure normal functioning. The relevant module, as showed in Figure \ref{fig:softwareresources}, is the \emph{$WSAN\_Cloud\_Interface$}. The ports attached to each block denote the flow of data among the parties: $wsan\_interface$ (WSANSoftware block) to $interface\_wsan$ ($WSAN\_Cloud\_Interface$ block), for instance. Finally, the block \emph{$Cloud\_Software$} has to conform to the selected approach (public or private cloud), represented by the package \emph{$Cloud\_Approaches$}.

\begin{figure}[!hbp]
	\centering
		\includegraphics{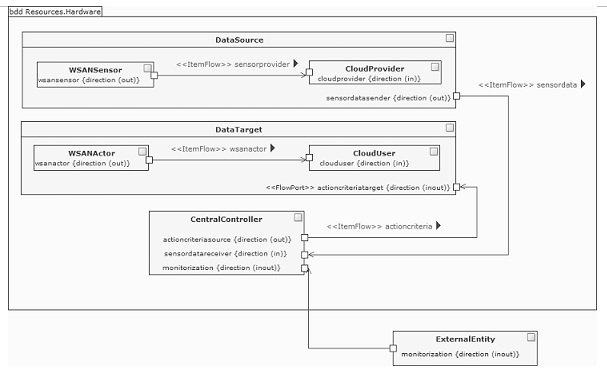}
	\caption{Block diagram about hardware resources (semi-automatic WSANs with the Cloud)}
	\label{fig:hardwareresources}
\end{figure}

	The block definition diagram sketched in Figure \ref{fig:hardwareresources} denote that the \emph{ExternalEntity} monitors the WSAN-Cloud integration through the \emph{CentralController} within the hardware architecture. Then, this controller issues commands to the \emph{DataTarget} block, where WSAN actors are related to the Cloud users. At the same time, the controller retrieves the information of the \emph{DataSource} block, where WSAN sensors are related to the Cloud providers, in order to process it and fulfill the subscriptions indicated by the external entity.

 In brief, integration of semi-automatic WSANs with the Cloud Computing can be performed associating the sensors to the providers, the actors to the users and monitoring these via a central controller. This schema is governed by a publish/subscribe mechanism which is monitored by an external entity (human user or another system).

\begin{figure}[!hbp]
	\centering
		\includegraphics{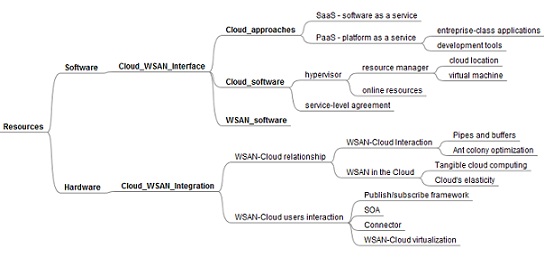}
	\caption{Resources for automatic WSANs integrated to the Cloud}
	\label{fig:automaticwsanresources}
\end{figure}

	\textbf{b. Set of specifications for the integration of automatic WSANs with the Cloud}: As mentioned earlier, in automatic WSANs sensors detect a phenomenon and send the data to actors which process the information to perform appropriate actions. In \cite{langendoerfer:2012} there is a comprehensive state of the art listing the approaches to integrate WSNs to the Cloud. Based on this work, Figure \ref{fig:automaticwsanresources} shows aspects linked to the integration of WSANs to the Cloud, where the data, instead of going from the Sensor to a Cloud Server, go directly to the defined actor(s). The integration of automatic WSANs to the Cloud has two main possibilities:
\begin{itemize}
	\item WSAN In The Cloud:The WSAN under study is part of the structure of the Cloud.
	\item WSAN With The Cloud: The WSAN under study interacts as a separate entity with the Cloud.
\end{itemize}

\begin{figure}[!hbp]
	\centering
		\includegraphics{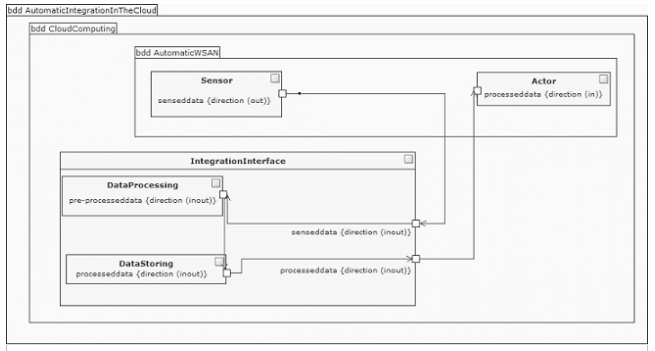}
	\caption{Automatic WSANs integrated to the Cloud as part of the latter one}
	\label{fig:wsaninthecloud}
\end{figure}

Figure \ref{fig:wsaninthecloud}	shows the schema of integration of automatic WSANs to the Cloud. In this scenario, the WSAN under study is considered as part of the structure of the Cloud. Sensors capture the data and send them to the \emph{IntegrationInterface} which uses the Cloud resources to filter the data, save them and then send the processed data to the Actors in order to execute the corresponding commands. In this proposal, data processing can take place in two parts: either in the interface or in the actor (it is assumed that the actor has a built-in processing unit); this depends on the model of integration adopted for the work. Moreover, external users can control the configuration of the integration through the interface.	
	
\begin{figure}[!hbp]
	\centering
		\includegraphics{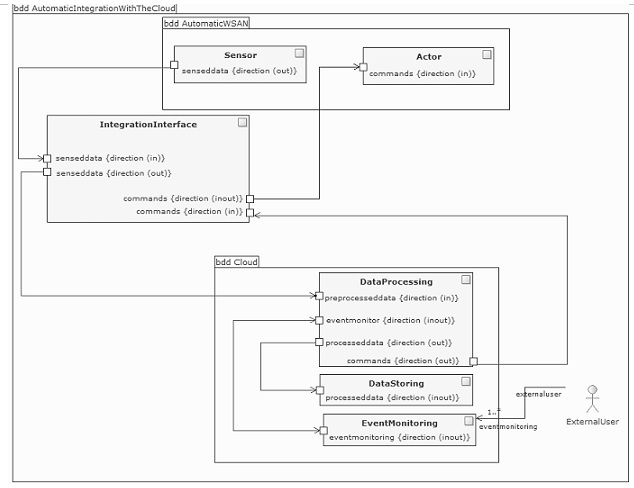}
	\caption{Automatic WSANs integrated to the Cloud as a separate entity}
	\label{fig:wsanwiththecloud}
\end{figure}

As mentioned before, WSANs can interact with the Cloud as a separate entity. Figure \ref{fig:wsanwiththecloud} indicates a proposal of how to integrate the two entities without establishing a tight structural relationship between them. The important concept is the introduction of the \emph{IntergrationInterface} as a mean to communicate the WSAN and the Cloud. A possible variation of this approach could be to directly connect the Cloud to the WSAN Actor to issue the commands to be executed. Finally, external users can monitor or control events through the Cloud. 
	
\begin{figure}[h]
	\centering
		\includegraphics{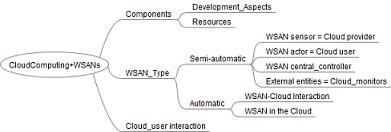}
	\caption{Summary of ideas for WSAN-Cloud integration}
	\label{fig:summaryforintegration}
\end{figure}

	After describing all the necessary elements and concepts to perform a WSAN-Cloud integration, Figure \ref{fig:summaryforintegration} summarizes the ideas to take into account. First of all, it is important to divide the work into two main Components: \emph{$Development\_Aspects$} and \emph{Resources}, where the first one studies the WSAN and the Cloud as separate entities while the second one assumes the common points for the integration. Then, the \emph{$WSAN\_Type$} defines the methods to be adopted for the integration. Finally, the \emph{$Cloud\_user interaction$} indicates the model to be used for the interaction of external users with the integration.
	
	\subsection{Application of the proposed design: a case study}
	A good way to demonstrate de validity of the proposed set of specifications for the integration of WSANs to the Cloud is to carry out a practical application. Following, a description of the context of study is presented and, then, the implementation of the proposal is fulfilled. The aim is to set the conditions of the work and, according to those plus the set of suggestions for integration, define an analysis-design document.
		
  \subsubsection{Description of the context of study}	
	The context of study is formed by a WSAN-based application with no integration to the Cloud. This application deals with the detection and extinction of forest fires. Sensors are deployed in a geographical area and are in charge of detecting temperature changes. Actors are responsible of extinguishing the fire of the area in danger \cite{kumar:2011}. 
				
	\begin{table}[h]
	\centering
	\begin{tabular}{|p{6cm}|p{10cm}|}
	\hline
	\textbf{Description} & \textbf{Value}  \\
	\hline
	Forest area division & $n X n$ square cell, called grids \\
	\hline
	Sensor node placement & center of each grid \\
	\hline
	Sensing range ($r$) and side of square cell ($D$) & $D = 2r$ \\
	\hline
	Node deployment & deterministic and static \\
	\hline
	Cluster Head (CH) node & includes nodes from a quadrant; placed at the corner of said quadrant \\
	\hline
	Actor nodes & contain fire extinguishing mechanism; are mobile; available at center of each quadrant \\
	\hline
	Fire propagation & uniform and constant speed \\
	\hline
	\end{tabular}
	\caption{Model assumptions for a WSAN application for forest fire detection and extinguishing \cite{kumar:2011}}
	\label{tab:modelassumptions}
\end{table}
		
\begin{figure}[ht]
	\begin{minipage}[b]{0.5\linewidth}
	\centering
	\includegraphics{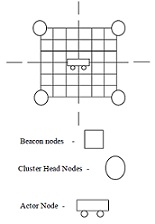}
	\caption{Quadrant model with different nodes \cite{kumar:2011}}
	\label{fig:figura8}
	\end{minipage}
	\hspace{0.5cm}
	\begin{minipage}[b]{0.5\linewidth}
	\centering
	\includegraphics{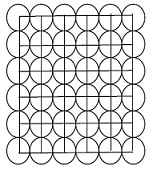}
	\caption{Sensor placement at the intersection of grid points \cite{kumar:2011}}
	\label{fig:figura9}
	\end{minipage}
\end{figure}

		For a better understanding, Figures \ref{fig:figura8} and \ref{fig:figura9} show the architecture of the nodes. Sensors are located in the center of each grid because it takes less nodes for an area, comparing to locating the nodes in each corner of the grid. Formally, sensors placed at the center of the grids require $n^2$ nodes, while sensors placed at the intersection of the grids requires $n^2 + 2n + 1$ nodes \cite{kumar:2011}.

	Regarding actor movement, ``the robot has to move in the prescribed environment'' \cite{kumar:2011}. For this case study, static path planning is applied because there are no moving objects and no obstacles. Table \ref{tab:formulas} includes the formulas to take into account for actor movement .
	
	\begin{table}[!hbp]
	\centering
	\begin{tabular}{|p{6cm}|p{7cm}|}
	\hline
	\textbf{Description} & \textbf{Formula}  \\
	\hline
	Distance for actor movement & $d = \sqrt{b^2+c^2}$ since $cos 90 = 0$ \\
	\hline
	Angle to reach target area & $\theta = \arctan(y/x)$ \\
	\hline
	\end{tabular}
	\caption{Formulas for a WSAN application for forest fire detection and extinguishing \cite{kumar:2011}}
	\label{tab:formulas}
\end{table}
	
	From the communications' viewpoint, it is assumed that sensors can connect and transmit data packets to the Cluster Head, without any inconvenience. Then, the Cluster Head processes the data, determines the quadrant from the location information and sends packets to the Actor \cite{kumar:2011}. There are two types of packets, the ones sent from the sensor to the Cluster Head (CH) or Beacon Node Packet \ref{tab:beacon}, Tables \ref{tab:beacon} and \ref{tab:cluster} respectively, and the ones sent from the CH to the Actor.
	
	\begin{table}[h]
	\centering
	\begin{tabular}{|p{2cm}|p{2cm}| p{2cm}|}
	\hline
	XC & YC & CHNO  \\
	\hline
	x Coordinate & y Coordinate & Cluster Head Number to which packet is to be sent. \\
	\hline
	\end{tabular}
	\caption{Format for the Beacon Node Packet \cite{kumar:2011}}
	\label{tab:beacon}
\end{table}

	\begin{table}[h]
	\centering
	\begin{tabular}{|p{2cm}|p{2cm}| p{2cm}| p{2cm}|}
	\hline
	XC & YC & QNO & AA \\
	\hline
	x Coordinate & y Coordinate & Quadrant Number & Address of the Actor to which the Packet is to be sent. \\	
	\hline
	\end{tabular}
	\caption{Format for the Cluster Head Node Packet \cite{kumar:2011}}
	\label{tab:cluster}
\end{table}
	
		In short, this WSAN-based application for detection and extinguishing forest fires is located in a $n X n$ squared area, where each square is a grid. Sensors are placed in the center of the grids, to reduce the number of deployed sensors. They are grouped in clusters, where each quadrant has a Cluster Head. The CH processes the information received and sends the commands to the corresponding actor in the quadrant \cite{kumar:2011}. 
		
	\subsubsection{Implementation of the proposed specifications to the described case study}
\begin{figure}[!hbp]
	\centering
		\includegraphics{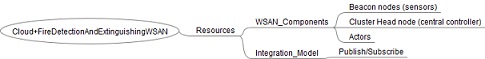}
	\caption{Schema for Cloud and Fire detection and extinguishing WSAN}
	\label{fig:schemaintegration}
\end{figure}

	This section is the practical application of the UML-based specifications for WSAN-Cloud integration. The WSAN described is a semi-automatic one because sensors, represented by the Beacon nodes, first transmit the sensed data to a central controller, characterized by the Cluster Head node, and then the latter one issues the commands to the corresponding actors.	Again, a Publish/Subscribe method is suggested for this application. In this approach, the monitoring of the WSAN will be performed via the Cloud (Figure \ref{fig:schemaintegration}). 
		
\begin{figure}[!hbp]
	\centering
		\includegraphics{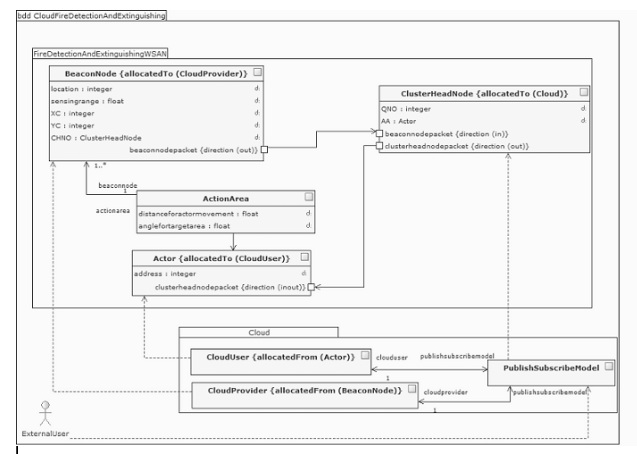}
	\caption{Block Diagram for Cloud and Fire detection and extinguishing WSAN}
	\label{fig:bddintegration}
\end{figure}

	Figure \ref{fig:bddintegration} portrays the design of the Fire Detection and Extinguishing WSAN integrated to the Cloud. The control of the WSAN is performed by an \textit{ExternalUser}, which might be a human user or any other entity. This control is carried out through the Publish/Subscribe model, which is allocated in the Cloud and has direct relationship to the Cloud users and providers. 

	On the other hand, the \textit{CloudProvider} block is allocated from the \textit{BeaconNode} block and the \textit{CloudUser} block is allocated from the \textit{BeaconNode} block. These two allocations imply a direct relationship between the WSAN under study and the Cloud. The \textit{ClusterHeadNode} receives the sensed data from the \textit{BeaconNode} and sends the command to the corresponding \textit{Actor}. The block \textit{ActionArea} comprises the necessary information about the geographical area where the actor will execute the received commands. It is important to note that the \textit{ClusterHeadNode} does not act independently from the Cloud, it is controlled by the \textit{Publish/SubscribeModel} block, which is inserted within the Cloud.

	The flow of data among the different blocks is spread through ports. For instance, when the \textit{BeaconNode} block transmits the sensed data, it does so through the port ``beaconodepacket'', with direction(out) towards the port in the \textit{ClusterHeadNode} also called ``beaconodepacket'', but with direction(in), which means it receives the sensed data. When the \textit{ClusterHeadNode} block issues the commands, it does so via the port ``clusterheadnodepacket'' with direction(out), heading to the port with the same name but located in the \textit{Actor} block and with direction(inout) which implies that the actor can both receive the commands and send its status to the ClusterHeadNode.

	Utilities like data storage and retrieval are managed via the Cloud. Data filtering is performed by the ClusterHeadNode, which in turn is under the control of the cloud. Besides, the cloud also controls the security, data access rules, and notification services via the Publish/Subscribe model inserted within it. Theoretically, the cloud is highly scalable, for this reason, the number of sensor nodes and actors can increase without affecting the performance of the integration. These utilities are part of the \textit{DevelopmentAspects} which considers the Cloud as a separate entity; for this reason, they are not explicitly indicated in the integration diagram.ls the security, data access rules, and notification services via the Publish/Subscribe model inserted within it. Theoretically, the cloud is highly scalable, for this reason, the number of sensor nodes and actors can increase without affecting the performance of the integration. These utilities are part of the \textit{DevelopmentAspects} which considers the Cloud as a separate entity; for this reason, they are not explicitly indicated in the integration diagram.
	
	\section{Conclusion and Future Work}
	The design of the integration of  Wireless Sensor and Actor Networks (WSANs) to the Cloud Computing is an emerging topic that should be expanded in order to provide the necessary mechanisms for a successful combination of the said entities. There are a number of aspects to consider for this integration. The work is divided into \textit{DevelopmentAspects} and \textit{Resources}, where the first one deals with each entity as a separate one in order to study its features and the latter one studies how the integration can be performed. Within the Resources, the type of WSAN influences on the overall architecture of the integration. Semi-automatic WSAN components are directly related to the Cloud actors and a model of integration has to be defined. Automatic WSANs can interact with the Cloud as a part of it or as a separate entity.
		
		This study applied several methodologies to achieve the proposed goals. In all cases, the analysis started with a mind map in order to show at a high level the issues involved in the integration. Then, SysML, as the proposed methodology by the OMG, contributed with two diagrams: Requirement Diagrams (RD) and Block Definition Diagrams (BDD). RDs were used to describe the conditions of the integration. BDDs included the components, their structure, their relationship and how the data flowed from one block to another. SysML is under development and for this reason the tools applied in this study are still in the early beginnings; for this reason, full compliance of the SysML specifications is not evident, reflected in the existence of explanations throughout this work.
		
		In general, this work aimed to offer some guidelines when designing the integration of WSANs to the Cloud Computing. The objectives were reached in their totality. However, there are several issues that need further research. For example, the models presented should be tested in more consistent designing tools, which should be ready in the short run. Moreover, a simulation of the integration is suggested to check the validity of this proposal.

\begin{singlespace}
	\bibliographystyle{plain}
	\bibliography{ElsevierManuscript} {} 
\end{singlespace}

\end{document}